\documentclass[screen]{acmart}

\usepackage{subcaption}
\usepackage{dirtytalk}
\usepackage[nameinlink]{cleveref}

\begin{document}

\copyrightyear{2026}
\acmYear{2026}
\setcopyright{cc}
\setcctype{by}
\acmConference[LAK 2026]{LAK26: 16th International Learning Analytics and Knowledge Conference}{April 27-May 01, 2026}{Bergen, Norway}
\acmBooktitle{LAK26: 16th International Learning Analytics and Knowledge Conference (LAK 2026), April 27-May 01, 2026, Bergen, Norway}
\acmDOI{10.1145/3785022.3785036}
\acmISBN{979-8-4007-2066-6/2026/04}

\title{Measuring the Impact of Student Gaming Behaviors on Learner Modeling}

\author{Qinyi Liu}
\orcid{0009-0003-4973-0901}
\affiliation{
    \department{Centre for the Science of Learning \& Technology (SLATE)}
    \institution{University of Bergen}
    \city{Bergen}
    \country{Norway}
}
\email{qinyi.liu@uib.no}

\author{Lin Li}
\orcid{0000-0002-4205-7975}
\affiliation{
    \department{Faculty of Information Technology}
    \institution{Monash University}
    \city{Melbourne}
    \country{Australia}
}
\email{lin.li@monash.edu}

\author{Valdemar Švábenský}
\orcid{0000-0001-8546-280X}
\affiliation{
    \department{Faculty of Informatics}
    \institution{Masaryk University}
    \city{Brno}
    \country{Czech Republic}
}
\email{valdemar@mail.muni.cz}

\author{Conrad Borchers}
\orcid{0000-0003-3437-8979}
\affiliation{
    \department{Human-Computer Interaction Institute, School of Computer Science}
    \institution{Carnegie Mellon University}
    \city{Pittsburgh, PA}
    \country{USA}
}
\email{cborcher@cs.cmu.edu}

\author{Mohammad Khalil}
\orcid{0000-0002-6860-4404}
\affiliation{
    \department{Centre for the Science of Learning \& Technology (SLATE)}
    \institution{University of Bergen}
    \city{Bergen}
    \country{Norway}
}
\email{mohammad.khalil@uib.no}

\begin{abstract}
The expansion of large-scale online education platforms has yielded vast amounts of student interaction data for knowledge tracing (KT). KT models estimate students’ concept mastery from interaction data, but the models' performance is sensitive to input data quality. Gaming behaviors, such as excessive hint use, may misrepresent students’ knowledge and undermine model reliability. However, systematic investigations of how different types of gaming behaviors affect KT remain scarce, and existing studies rely on costly manual analysis that does not capture behavioral diversity. In this study, we conceptualize gaming behaviors as a form of data poisoning, defined as the deliberate submission of incorrect or misleading interaction data to corrupt a model’s learning process. We design Data Poisoning Attacks (DPA) to simulate diverse gaming patterns and systematically evaluate their impact on KT model performance. Moreover, drawing on advances in DPA detection, we explore unsupervised approaches to enhance the generalizability of gaming behavior detection. We find that KT models performance tend to decrease especially for random guess behaviors. Our findings provide insights into the vulnerabilities of KT models and highlight the potential of adversarial methods for improving the robustness of learning analytics systems.
\end{abstract}



\maketitle

\section{Introduction}
\label{intro}

With the proliferation of online education platforms such as Massive Open Online Courses, Learning Management Systems, and Intelligent Tutoring Systems, vast amounts of student interaction data are being recorded \cite{shen2024survey}. These data, which encompass students’ responses, study duration, and interaction patterns, form the foundation for Adaptive Learning Systems (ALS) \cite{dai2021knowledge}. Knowledge Tracing (KT) is a core modeling technique for ALS, dynamically tracking students’ knowledge states by modeling their concept mastery based on graded attempts at problem steps in log data~\cite{corbett1994knowledge}. Leveraging KT, ALS can select appropriate learning materials, thereby constructing personalized learning pathways. However, the reliability and effectiveness of learning analytics predictions hinge on data quality. When data quality degrades or becomes contaminated, the performance of learning analytics models may be significantly impaired \cite{Li2019OptOut}. Studies have shown that label errors or anomalous response sequences can notably affect the Area Under the Curve (AUC) (a widely used metric in machine learning and statistics to evaluate the performance of a binary classification model~\cite{handbook-la2022}) and knowledge state estimation of KT models \cite{Wang2025KnowledgeTracing}.

In real-world settings, ensuring the quality of student data used by ALS is challenging. Degradation of data quality can stem from external factors or student behaviors such as gaming the system, where students bypass cognitive effort through trial-and-error, hint abuse, or rapid clicking to “game” the system for correct answers~\cite{baker2008gaming}. Such behaviors generate data inconsistent with students’ true knowledge levels, violating the core assumption of KT models—that student responses accurately reflect their knowledge states and improve with practice over time \cite{koedinger2023astonishing}. Previous research has mainly addressed the motivations, patterns, and detection of gaming behaviors \cite{baker2008gaming, paquette2019comparing}, yet analyses of different types of gaming behaviors and their impact on the KT model are lacking. The reasons for this gap are outlined below. 

One major challenge in understanding the impact of gaming behaviors on KT-based learning systems lies in the annotation of such behaviors. First, identifying and labeling such behaviors typically relies on expert manual analysis, which is time-consuming and costly \cite{paquette2019comparing}. Second, gaming behaviors are highly heterogeneous; for instance, some students avoid thinking by repeatedly requesting hints, while others try to attempt correct answers through rapid random attempts \cite{paquette2019comparing}. To address this need for systematic and controllable manipulation, we propose conceptualizing gaming the system behaviors as “data contamination,” noting their similarity to patterns in Data Poisoning Attacks (DPA) in general machine learning, such as injecting erroneous labels or anomalous interactions to disrupt model learning. Thus, in the absence of diverse, high-quality labeled data, we employ DPA to simulate different types of gaming behaviors, systematically examining their potential impact on KT model performance.

Furthermore, detecting gaming behaviors is similarly constrained by reliance on manual annotations for detector training, limiting the generalizability of existing detectors to unseen samples \cite{Levin2022GamingDetector}. In the field of machine learning security, detection methods for DPA are well-established, including outlier detection, gradient- or loss-based analysis, and influence-based approaches, which demonstrate strong generalization to unseen attack samples \cite{Zhao2025DataPoisoningSurvey}. We propose that these detection strategies can inspire gaming behavior detectors, for example by using principle-based simulation of gaming or unsupervised anomaly detection to reduce reliance on manual labels. To the best of our knowledge, this unique approach has not yet been studied in learning analytics contexts. The contributions of this paper are:

\begin{itemize}
    \item We customize DPA to simulate student gaming behaviors in ALS based on different behavioral patterns.
    \item We compare the impact of varying types and proportions of gaming behaviors on KT model performance.
    \item We identify directions for future KT model design to address gaming behaviors and improve robustness.
    \item We discuss how DPA detection methods can inform the development of traditional gaming behavior detectors.
\end{itemize}

\section{Background, Motivation, and Problem Statement}

The background of our study combines three parts: knowledge tracing (\Cref{subsec:related-work-KT}), gaming behaviors (\Cref{subsec:related-work-gaming}), and data poisoning attacks (\Cref{subsec:related-work-DPA}). Subsequently, 
\Cref{subsec:RQ} summarizes the motivation and proposes research questions based on the gaps identified in the prior work.

\subsection{Knowledge Tracing}
\label{subsec:related-work-KT}

KT is an essential aspect of ALS, as recognized by the LAK conference community~\cite{Effenberger2020afm}. KT is a key student modeling technique that has been widely applied in ALS to dynamically assess students’ knowledge states and support personalized learning \cite{corbett1994knowledge}. Specifically, the KT task can be formulated as a supervised learning problem: given a student’s historical interaction sequence up to time $t$, denoted as $X_T = (x_1, x_2, \dots, x_t)$, the goal is to predict the student’s performance on the next interaction $x_{t+1}$. Each interaction $x_t = (q_t, a_t)$ is defined as a tuple, where $q_t$ denotes the knowledge component (KC), ID of the attempted question at the time step $t$, and a binary label $a_t$ indicating whether the response was correct. KT models aim to estimate the probability of a correct response at the next time step, i.e., $p(a_{t+1}=1 \mid q_{t+1}, X_T)$ \cite{dai2021knowledge}. The knowledge states measured by KT can further be leveraged within learning analytics to personalize students’ learning pathways, thereby enhancing learning efficiency in ALS \cite{abdelrahman2023knowledge}.

KT models can be broadly categorized into three groups, each rooted in artificial intelligence techniques \cite{dai2021knowledge}. First, probabilistic graph models use dynamic Bayesian networks to infer probability distributions representing students’ knowledge states. A well-known example is Bayesian Knowledge Tracing (BKT) \cite{corbett1994knowledge}. Second, logistic regression models, such as the Additive Factors Model (AFM), predict the probability of a correct response on a given knowledge component through a regression equation \cite{Pelanek2017}. Third, deep learning models, such as Deep Knowledge Tracing (DKT), leverage neural networks to capture complex temporal dependencies in student interactions, thereby improving predictive accuracy~\cite{piech2015deep}. Despite the diverse nature of KT models, they all rely heavily on the quality of input data to accurately model knowledge states. Studies have demonstrated that KT, whether based on Bayesian, logistic model, or deep learning approaches, is highly sensitive to input data quality \cite{Xiong2016GoingDeeper, Yang2025DifficultyAware}. Data format, noise, lack of diversity, and incomplete labeling can all impair the model's ability to accurately track a student's knowledge state. However, it remains unclear whether noise or bias introduced by student behavior, such as random guessing and other gaming behaviors \cite{baker2008gaming}, poses a threat to KT models, and, consequently, hinders ALS from providing a better learning experience.

\subsection{Gaming Behaviors}
\label{subsec:related-work-gaming}

\say{Gaming the system} refers to student behaviors that exploit the mechanics of educational software in order to advance without engaging meaningfully with the underlying instructional content~\cite{baker2004off}. Examples include rapid guessing, systematic answer trialing, and excessive hint requests (often until a bottom-out hint with the solution is provided). Such actions decouple observable progress within the system from genuine learning, thereby \say{poisoning} the data used to build and evaluate KT models. Gaming behaviors is a phenomenon that has been widely observed across several ALS~\cite{baker2004off,gong2010impact,Huang2023GamingDetection}, though the specific frequency differs by how complex and difficult instruction of each system is \cite{baker2009educational}.

Prior work has demonstrated that gaming behaviors impact estimates of student learning. For instance, Gong et al.~\cite{gong2010impact} showed that students who engage in gaming exhibit flat learning parameters (i.e., minimal or no improvement in knowledge acquisition over time), which is uncommon for tutoring systems \cite{koedinger2023astonishing}. Similarly, task-irrelevant on-system behaviors—sometimes described as WTF (\say{Without Thinking Fastidiously}) behavior \cite{wixon2012wtf}—further reduce the effectiveness of instructional feedback. Unlike traditional off-task behavior, which occurs outside the system (e.g., chatting with peers), WTF behavior manifests through playful or irrelevant uses of the interface, such as inserting nonsensical inputs, dragging irrelevant objects into simulations, or drawing shapes with graphing tools. Although these behaviors occur within the boundaries of the system, they are disconnected from the intended learning goals.

The presence of such behaviors raises serious challenges for learner modeling and learning analytics. Models such as BKT or AFM assume that correct and incorrect responses provide evidence of learning over time, and typically estimate that students improve as a function of the amount of practice they engage in \cite{koedinger2023astonishing}. When data are contaminated by gaming or WTF behavior, these assumptions may no longer hold, leading to overestimates of instructional effectiveness or unreliable conclusions about learning. In practice, this means that datasets with higher proportions of gaming behaviors might yield fewer meaningful insights into learning processes, though we are not aware of any prior research that has systematically quantified this impact through simulation.

Detection of gaming behaviors has been an active research area, and many effective, reliable, and valid detectors have been developed~\cite{paquette2014reengineering,Huang2023GamingDetection}. Gaming detectors can be mainly categorized into knowledge-engineering-based models, machine-learning-based models, and hybrid models that combine the two approaches~\cite{Huang2023GamingDetection}. Both types rely on human labeling of log data sequences or classroom observations to establish ground truth labels for developing gaming detectors \cite{ocumpaugh2015baker}. Other times, the evaluation process involves two or more human coders providing labels, given sufficient inter-rater reliability~\cite{Huang2023GamingDetection}. 
Features of student behavior from learning system log data (e.g., hint usage patterns) are then used to detect whether a student is gaming or not, either using expert-based decision rule models \cite{paquette2014reengineering} or supervised machine learning \cite{paquette2018system,paquette2019comparing}. 
However, ensuring consistency of gaming detection results across different samples and achieving generalization to new samples remain ongoing challenges in the field \cite{baker2008gaming, paquette2014disengagement, Huang2023GamingDetection}.

Here, we draw on insights from data patterns identified in prior work as predictive of gaming behaviors to inform the design of data poisoning attacks that simulate these behaviors and contaminate training data. Specifically, to generate gaming behaviors, we employ rules derived from latency thresholds and action sequences (e.g., short hint response times, \say{guess–read–guess} patterns). For WTF behavior, distinct detection logic is required, as its patterns differ substantially from strategic exploitation of the system. Given the empirical evidence that gaming and WTF behavior are associated with less learning~\cite{gong2010impact}, the novelty of our study lies in using existing methods to generate \say{poisoned} data that must be accounted for when analyzing student interactions in adaptive learning systems. This enables researchers and practitioners to better estimate the impact of such behavior on student models via simulation.

\subsection{Data Poisoning Attacks}
\label{subsec:related-work-DPA}

Data Poisoning Attacks (DPA) are a class of adversarial attacks in machine learning, aiming to deliberately inject malicious or erroneous data into training datasets to degrade model performance~\cite{Fan2022SurveyDataPoisoning}. Such adversarial attacks are important for enhancing the robustness, security, and reliability of models. First, DPA can expose vulnerabilities of a model by introducing malicious samples, helping researchers identify potential flaws and improve model design~\cite{AlMaliki2023AdversarialMachineLearning}. Second, by studying the effects of DPA, researchers can develop more robust models or defense mechanisms to mitigate potential malicious behaviors \cite{AlMaliki2023AdversarialMachineLearning}. Third, DPA can simulate real-world adversarial scenarios, allowing models to be tested under hostile conditions and evaluate their security and robustness \cite{Fan2022SurveyDataPoisoning}. This is especially common in the autonomous driving domain; for example, \citet{Gu2017BadNets} used backdoor-based DPA to simulate situations where stickers are placed on road signs to mislead recognition systems. Overall, DPA aims to improve a model's generalization ability and robustness by simulating worst-case scenarios. However, predictive models in the social sciences have faced substantial challenges in realizing generalizability and robustness \cite{yarkoni2022generalizability}. This is why, in the field of adversarial machine learning, researchers have long advocated incorporating adversarial attacks such as DPA into social science models to enhance their robustness \cite{AlMaliki2023AdversarialMachineLearning}. As a field related to social sciences, learning analytics holds significant potential for applying DPA. Since learning analytics typically relies on large-scale student behavioral data for predictive modeling, data poisoning can lead to misguided interventions, fairness issues, and ethical concerns. By integrating DPA testing during the development of learning analytics models, it is possible to enhance model iteration, thereby improving robustness and generalizability in ALS and reduce the potentially harmful impact on students learning.

Research on DPA detection is also fruitful \cite{Zhao2025DataPoisoningSurvey}. Notably, many commonly used DPA detection methods are based on unsupervised learning. For example, detection methods based on statistical anomalies identify outliers by assuming the distributional characteristics of a clean dataset, without requiring prior labeling of poisoned data. \citet{Kure2025DetectingPreventing} employed statistical anomaly detection to effectively mitigate the impact of data poisoning, thereby improving the robustness of the model and restoring accuracy by 15–20\%. In addition, influence functions serve as another effective approach to quantify the impact of individual data points on the model’s output, making it possible to identify highly influential points—i.e., poisoned data~\cite{Zhao2025DataPoisoningSurvey}. This method likewise does not require labeled data.

In ALS, gaming the system behaviors, such as random guessing, rapid answering, or deliberately incorrect responses, are common forms of malicious behavior that introduce noise or biases into interaction data, deviating from normal learning patterns \cite{baker2008gaming}, and potentially degrading the performance of KT models. These behaviors exhibit significant similarities to DPA in their effects, as both disrupt the model’s learning process by injecting low-quality or misleading data. Specifically, abnormal response patterns generated by gaming behaviors (e.g., incorrect labels caused by random guesses) resemble the random or targeted erroneous data injected in DPA, making DPA an ideal tool for simulating such behaviors. As mentioned in \Cref{intro}, obtaining real-world labeled gaming behaviors data is often challenging due to data scarcity or the high cost of manual annotation \cite{baker2008gaming, Levin2022GamingDetector}. In contrast, DPA can generate controllable and scalable simulated data, effectively addressing this limitation. DPA can simulate extreme behavioral scenarios (such as large-scale random guessing or deliberate incorrect responses), allowing KT models to be tested under stress conditions and providing insights for optimizing recommendation algorithms in ALS.

To the best of our knowledge, no prior studies in learning analytics have systematically simulated gaming the system behaviors at varying scales and types. Previous studies focused on the detection and impacts of gaming behaviors~\cite{gong2010impact}. DPA provides a systematic approach to such simulations, addressing a gap at the intersection of learning analytics and machine learning security. By investigating the impact of DPA on ALS, researchers can guide system designers in developing more robust models, for instance, by identifying learning path deviations caused by gaming behaviors and thereby optimizing personalized learning experiences. This effort to introduce adversarial attacks into the educational domain aligns with scholarly calls to apply DPA in social science models to enhance their robustness \cite{AlMaliki2023AdversarialMachineLearning}.

\subsection{Motivation and Research Questions}
\label{subsec:RQ}

From the above literature review, it is evident that KT models are highly sensitive to data quality \cite{Xiong2016GoingDeeper}. Gaming behaviors in ALS, including random guessing or rapid answering, introduce low-quality data that resemble DPA in adversarial machine learning, potentially misleading KT models and reducing the effectiveness of personalized learning. To date, the specific impact of gaming behaviors on KT models has not been thoroughly investigated, and the high cost of obtaining real-world gaming behaviors data has limited related research \cite{baker2008gaming, Levin2022GamingDetector}. DPA offer a controllable and systematic approach to simulate large-scale or diverse gaming behaviors, test the robustness of ALS, and provide new perspectives for detection methods. Based on this, the present study raises the following research questions (RQs):  
\begin{enumerate}
    \item[RQ1:] How can data poisoning attacks be leveraged to simulate students’ gaming behaviors in ALS?
    \item[RQ2:] Do different types of simulated gaming behaviors have distinct effects on KT models performance?
    \item[RQ3:] What insights can DPA detection methods provide for traditional gaming behaviors detection?
\end{enumerate}

\section{Research Methods and Implementation of Experiments}

This section first outlines the datasets, selected KT models, and implementation details. It then describes the experiments addressing RQ1 and RQ2. RQ3 is examined qualitatively by combining these results with prior work on DPA detection, highlighting vulnerabilities of KT models under data poisoning. Detailed insights on their potential for enhancing gaming behavior detection are discussed in \Cref{sec:discussion}. The research code is publicly available at \cite{supplementary-materials}.

\subsection{Datasets}

Our study uses two baseline datasets, labeled A and B, which we detail below. The datasets were collected within well-established ALS. We selected these datasets because they include detailed student activity and performance records comprising gaming behaviors, enabling us to address our research questions.

Dataset A is a publicly available dataset released for the 2017 ASSISTment Longitudinal Data Mining Competition\footnote{\url{https://sites.google.com/view/assistmentsdatamining/}}, comprising 942,752 records from 1,709 middle school students in United States~\cite{Patikorn2020datasetASSISTments}. These records were collected from students' interactions with the ASSISTments system while practicing problems covering 101 middle-school mathematical concepts during 2004–2006. With its large-scale student interactions and skill coverage and widespread use as a benchmark dataset for comparing different KT models, this dataset is ideal for our study.

Dataset B originates from the study by~\citet{Baker2008datasetHamptonAlg} and was used in more recent learning analytics studies, such as~\cite{Levin2022GamingDetector}. This dataset was provided to us by the study authors. The dataset contains 240,237 actions performed by 59 students using the Cognitive Tutor system to learn 87 unique skills within the domain of algebra. The actions were collected from high school students in the USA~\cite{Baker2008datasetHamptonAlg}. Compared to Dataset A, this dataset includes fewer students, but it is longitudinal (collected throughout an entire school year).

Both datasets have an equivalent set of features. For our modeling purposes using KT, the relevant features are: student ID, skill name, and answer correctness, which is binary coded.

\subsection{KT Models}

In \Cref{subsec:related-work-KT}, KT models were categorized into three groups: probabilistic graph models, logistic regression models, and deep learning models \cite{dai2021knowledge}. To investigate the impact of gaming behaviors on KT performance, we select one representative method from each category: BKT from probabilistic graph models \cite{corbett1994knowledge}, AFM from logistic regression models \cite{Pelanek2017}, and DKT from deep learning models \cite{piech2015deep}. These methods are widely adopted in learning analytics, as evidenced by their use in LAK publications~\cite{Effenberger2020afm, Long2018afm, goutte2020confident}. Due to space limitations, please refer to \cite{supplementary-materials} for the complete mathematical definition for the KT models. We implemented these three models (BKT, AFM, and DKT) using the \texttt{sklearn} module in Python 3. For each model, a single script was designed to uniformly process both our datasets. 

\subsubsection{BKT} 

Widely acknowledged as a canonical probabilistic method for modeling learners' knowledge states, BKT treats students' mastery of a knowledge component or skill as a latent variable that can be inferred from binary responses (i.e., correct or incorrect) to items aligned with that skill~\cite{corbett1994knowledge}. In its standard form, BKT estimates four core parameters from students' sequential responses on a set of items associated with that skill, including: $P(L_0)$, the prior probability of a skill initially known; $P(T)$, the probability of transitioning from not-known to known after each response; $P(S)$, the (slip) probability of answering an item incorrectly despite knowing the skill; and $P(G)$, the (guess) probability of answering an item correctly despite not knowing the skill. Given these four parameters, the probability of a correct response at step $t$ is the sum of the probability of knowing the skill and not slipping and the probability of not knowing the skill and guessing correctly, where the probability of knowledge state at current step (i.e., knowing a skill or not) is updated from the previous knowledge state via the transitioning probability $P(T)$. It is important to note that BKT requires sequentially ordered attempts for each skill, as the model explicitly updates mastery over time. We used the \texttt{pyBKT} library\footnote{\url{https://github.com/CAHLR/pyBKT}}~\cite{badrinath2021pybkt} to fit the standard BKT, which includes several variants and follows \texttt{sklearn} conventions.

\subsubsection{AFM} 

In simple terms, AFM is a logistic regression model, which predicts the probability that a student will correctly answer a learning item. The exact definition can be found in~\cite{Effenberger2020afm}. To fit the model, the corresponding dataset must uniquely identify students and learning items (also called knowledge components), which represent skills. The dataset also needs to label binary correctness for each student-item pair, i.e., whether the student answered the item correctly or incorrectly. An important note is that AFM is designed to model learning over repeated opportunities, i.e., it considers how many times a student has attempted a particular skill before a current attempt.

We implemented AFM using the \texttt{LogisticRegression} class from \texttt{sklearn}. Most hyperparameters were set to defaults as recommended by the module documentation.\footnote{\url{https://scikit-learn.org/stable/modules/generated/sklearn.linear_model.LogisticRegression.html}} We changed the default solver to \texttt{saga} as it handles larger datasets more quickly. In order to ensure convergence with \texttt{saga}, we preprocessed the features to be on the same scale, as recommended by the documentation. Finally, we experimentally set the maximum number of iterations to 100. With larger values, the models stopped improving notably, and convergence was much slower.

\subsubsection{DKT} 
Unlike traditional KT models that summarize the knowledge state as a single latent variable per skill, DKT considers knowledge as a high-dimensional dynamic state, enabling richer properties and cross-skill dependencies to be modeled~\cite{piech2015deep}. DKT applies a recurrent neural network to the sequence of learner-item interactions. At each time step, the input encodes the practiced skill tested through the current item and whether the response was correct, and the network outputs a vector of probabilities of answering correctly at next step (a distribution over skills), trained against the next observed interaction. In empirical studies, DKT has served as a strong baseline and often outperformed classical KT variants. We train and evaluate DKT using the TensorFlow implementation provided in \cite{yeung2018addressing,yeung2019incorporating}\footnote{\url{https://github.com/ckyeungac/ADM2017}}.

\subsection{RQ1 -- Simulating Gaming Behaviors via DPA in ALS}

To answer RQ1, this section describes how we use different DPA to simulate students’ gaming behaviors in ALS, including varying poisoning ratios of 5\%, 25\%, and 50\% for each attack type to mimic scalable and extreme scenarios (see \Cref{tab:dpa_gaming_mapping}).

\begin{table*}[h]
\centering
\caption{Mapping of DPA strategies to gaming behaviors.}
\small
\label{tab:dpa_gaming_mapping}
\begin{tabular}{|l|l|p{3.7cm}|}
\hline
\textbf{DPA strategy} & \textbf{How to poison} & \textbf{Gaming the system behavior} \\
\hline
Random Error Attack & 
\begin{tabular}[t]{@{}l@{}}
1) Flip correctness labels \\ 
2) Set response time to 1 second
\end{tabular} & Random guessing, WTF \\
\hline
Hint Abuse Attack & 
\begin{tabular}[t]{@{}l@{}}
1) Insert multiple hint requests with short intervals \\ 
2) Submit answers quickly after hints \\ 
3) Some records only request hints without submitting answers
\end{tabular} & Hint abuse \\
\hline
Sequential Pattern Attack & 
\begin{tabular}[t]{@{}l@{}}
1) Generate repeated submissions with short response times \\ 
2) Early answers incorrect, final answer correct
\end{tabular} & Systematic Answer Trialing, Guess-Read-Guess \\
\hline
\end{tabular}
\end{table*}

\textbf{Random error attacks.} To simulate gaming behaviors such as random guessing and WTF, we designed a random error attack. Random guessing refers to students attempting to guess the correct answer by making multiple rapid incorrect attempts without considering the problem content or engaging in knowledge-based reasoning \cite{vanacore2017effect}. WTF exhibits similar behavioral patterns but differs in purpose. As described partly in \Cref{subsec:related-work-gaming}, WTF emphasizes careless interactions without thoughtful consideration, which derail the learning process. The common characteristics of both behaviors are extremely short response times and high error rates \cite{wixon2012wtf}.

These behaviors are highly analogous to label flipping in DPA, where labels are deliberately altered (e.g., flipping correct labels to incorrect ones) \cite{Fan2022SurveyDataPoisoning}. Based on the behavioral patterns of random guessing and WTF, and inspired by the concept of label flipping, we designed the random error attack. In this attack, correct answers are flipped into incorrect ones, and the response time is set to one second. The choice of one second is motivated by prior research~\cite{WiseKong2005ResponseTimeEffort, vanacore2017effect}, which shows that response times under five seconds are strongly associated with guessing, while for problems that do not require substantial reading, guessing behaviors may even occur with response times below three seconds. Considering that the dataset we plan to use primarily consists of mathematics problems that require little reading support (see below), we fixed the response time at one second.

\textbf{Hint Abuse Attack.} To simulate hint abuse behavior in ALS, we designed the hint abuse attack. Hint abuse refers to students rapidly and repeatedly requesting system hints to directly obtain answers, rather than using hints as learning aids. \citet{baker2008gaming} noted that students may request hints in quick succession, skipping intermediate hints to obtain the \say{bottom-out hint,} which directly reveals the answer, thereby circumventing the learning process. A key characteristic of this behavior is that students make multiple hint requests within a few seconds (typically 0.5–1 second intervals), ignore the problem content, and either copy the answer to progress quickly or browse hints without submitting answers to bypass learning. \citet{paquette2014disengagement} and \citet{vanacore2017effect} further support this observation, stating that fast responses ($<3$ seconds) are often associated with low cognitive effort, such as exploiting hints to obtain answers.

To reproduce this behavior, we applied the following settings:
\begin{enumerate} 
    \item Insert 3--5 hint request records (\say{hint\_request}) into the data for randomly selected students and questions, with intervals of 0.5--1 second.  
    \item After the final hint request, simulate the submission of a correct answer with a response time of 0.5 seconds, representing rapid copying behavior.  
    \item For 50\% of the hint request records, omit corresponding answer submissions, reflecting students who skip intermediate hints and seek only the bottom-out hint without meaningful engagement.  
\end{enumerate}

This design can mislead knowledge tracing models into overestimating student ability, thereby testing the robustness of adaptive learning systems against hint abuse behavior.

\textbf{Sequential Pattern Attack.} To simulate systematic answer trialing or guess-and-check in gaming behaviors, we designed the Sequential Pattern Attack. These behaviors describe a structured pattern of answer attempts, where students systematically submit different answers and use system feedback to gradually identify the correct option, rather than relying on genuine understanding~\cite{baker2004off}. For example, a student may repeatedly enter variant answers in quick succession, immediately changing their response after an error, until success.

To reproduce this behavior, we incorporated characteristics of the label flipping attack in data poisoning and applied the following settings:
\begin{enumerate} 
    \item The answer sequence is arranged in option order (e.g., first attempt = A, second = B, third = correct answer C). 
    \item Each submission is assigned a short response time ($<2$ seconds) to reflect rapid guessing. 
    \item Correctness is manipulated by flipping labels: earlier submissions are marked as incorrect, while the final submission is set as correct. 
\end{enumerate}

\subsection{RQ2 -- Comparing the Effects of Gaming on KT Performance}

\begin{figure*}[h]
    \centering
    \includegraphics[width=0.8\textwidth]{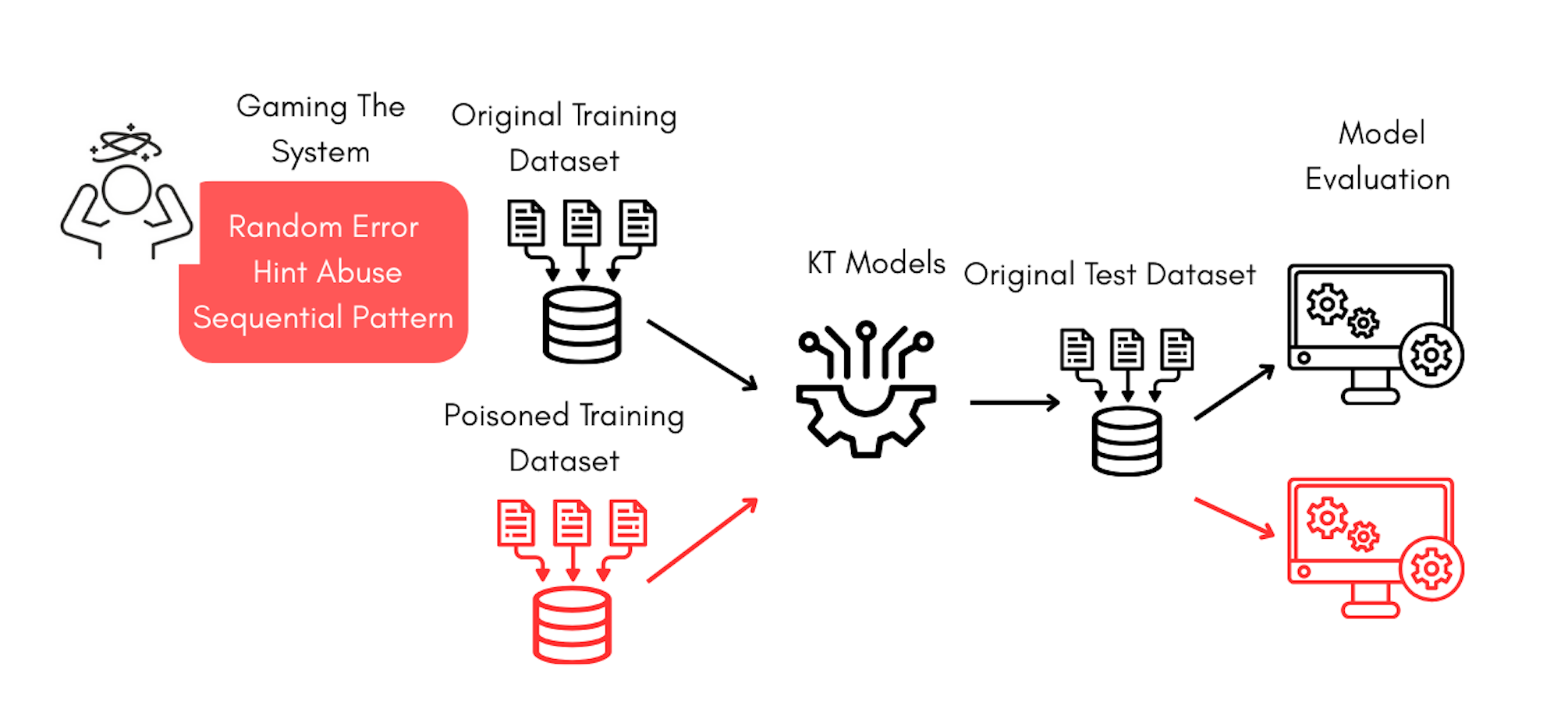}
    \caption{Experimental workflow for simulating gaming behaviors in adaptive learning systems.}
    \label{fig1}
\Description{A diagram providing an overview of the study design.}
\end{figure*}

To address RQ2, we designed the following procedure (see \Cref{fig1}): the original dataset is divided into training and testing sets at an 80:20 ratio. The training set is injected with simulated gaming data (including Random Error, Hint Abuse, and Sequential Pattern) to generate poisoned training datasets, while the testing set is untainted. This design aims to evaluate the impact of simulated data by training KT models on poisoned training data and evaluating them on original testing data, validating the synthetic data generators and enabling comparison of the effects of different attack types on model performance \cite{Liu2024}. Following the training phase, we assessed the models' performance using the AUC~\cite{handbook-la2022}, a standard metric widely adopted at LAK to evaluate various learning analytics models \cite{Long2018afm, wixon2012wtf}. Finally, to examine how model performance varies across different poisoning strategies, we trained the models on several poisoned variants of the original dataset, aligning with the model evaluation step outlined in the experimental workflow.

To ensure robust evaluation and prevent data leakage between training and test sets, we employed two distinct and data-splitting strategies: \textit{student-level} and \textit{skill-level} splits. Skill-level splits group data by skills to address the sensitivity of KT models in unseen skills, while student-level splits partition data by learner identifiers to simulate training on one cohort and deploying on another. Together, these approaches assess both inter-student generalization and skill-specific vulnerabilities in ALS, while preventing data leakage and aligning with real-world contexts.

\par For the skill-level split, we ensured that the training set included all skills appearing in the test set, since some KT models (e.g., BKT, AFM) are sensitive to unseen skills \cite{corbett1994knowledge}. This ensures that model performance is not adversely affected by skills absent from the training data. For the train–test split, we first grouped each student’s sequences by skill/knowledge, then, for each skill, we randomly sampled 20\% of the students for the test set, while the remaining 80\% were included in the training set, ensuring the model is always evaluated on students per skill it has not seen during training. This split is critical for evaluating model performance on skill mastery under DPA, particularly for targeted attacks on specific skills, and provides insights into skill-specific vulnerabilities. 

For the student-level split, we partitioned the data by student identifiers, ensuring no overlap between training and test sets. This approach mirrors practical scenarios where a model is trained on one cohort of learners and applied to a future cohort, often encountering the same or overlapping skills, which is typical in ALS~\cite{baker2008gaming}. This split also aligns with the design of DPA, which simulate gaming behaviors (e.g., random guessing, hint abuse) specific to individual students, allowing us to assess model robustness against such behaviors. To account for the sensitivity of certain KT models to unseen skills, the student-level split similarly ensured that all skills in the test set were included in the training set. 

\section{Results}

 This section presents the results of our experiments (addressing both RQ1 and RQ2), showcasing the impact of DPA on KT models through percentage change in AUC. (RQ3 is answered in \Cref{sec:discussion}.) The percentage change is calculated as
\[
\frac{AUC_{\text{poisoned}} - AUC_{\text{original}}}{AUC_{\text{original}}} \times 100,
\]
and is visualized in bar charts comparing skill-level and student-level splits across datasets and attack types. The details of the experiment results can be seen in the supplementary materials~\cite{supplementary-materials}.

\textbf{Overall Comparison Across DPA Types.} Across all experiments, Random Error Attacks consistently caused the most severe performance degradation (average AUC drop of $-15.2\%$ across models and datasets), followed by Hint Abuse Attacks ($-1.83\%$ on Dataset A), while Sequential Pattern Attacks had the least impact (average change near 0\%, with some slight positive gains). This suggests that unstructured noise (Random Error) is more disruptive to KT models than patterned or context-specific attacks. Regarding poisoning ratios, performance degradation generally increased with higher ratios (e.g., 50\% ratio led to 2--3${\times}$ larger drops than 5\% across types), except for Sequential Pattern Attacks, where changes remained minimal regardless of ratio.

\begin{figure*}[htbp]
    \centering
    \captionsetup{font=normalsize} 

    \begin{subfigure}[b]{0.49\textwidth}
        \centering
        \resizebox{1.1\textwidth}{!}{\includegraphics{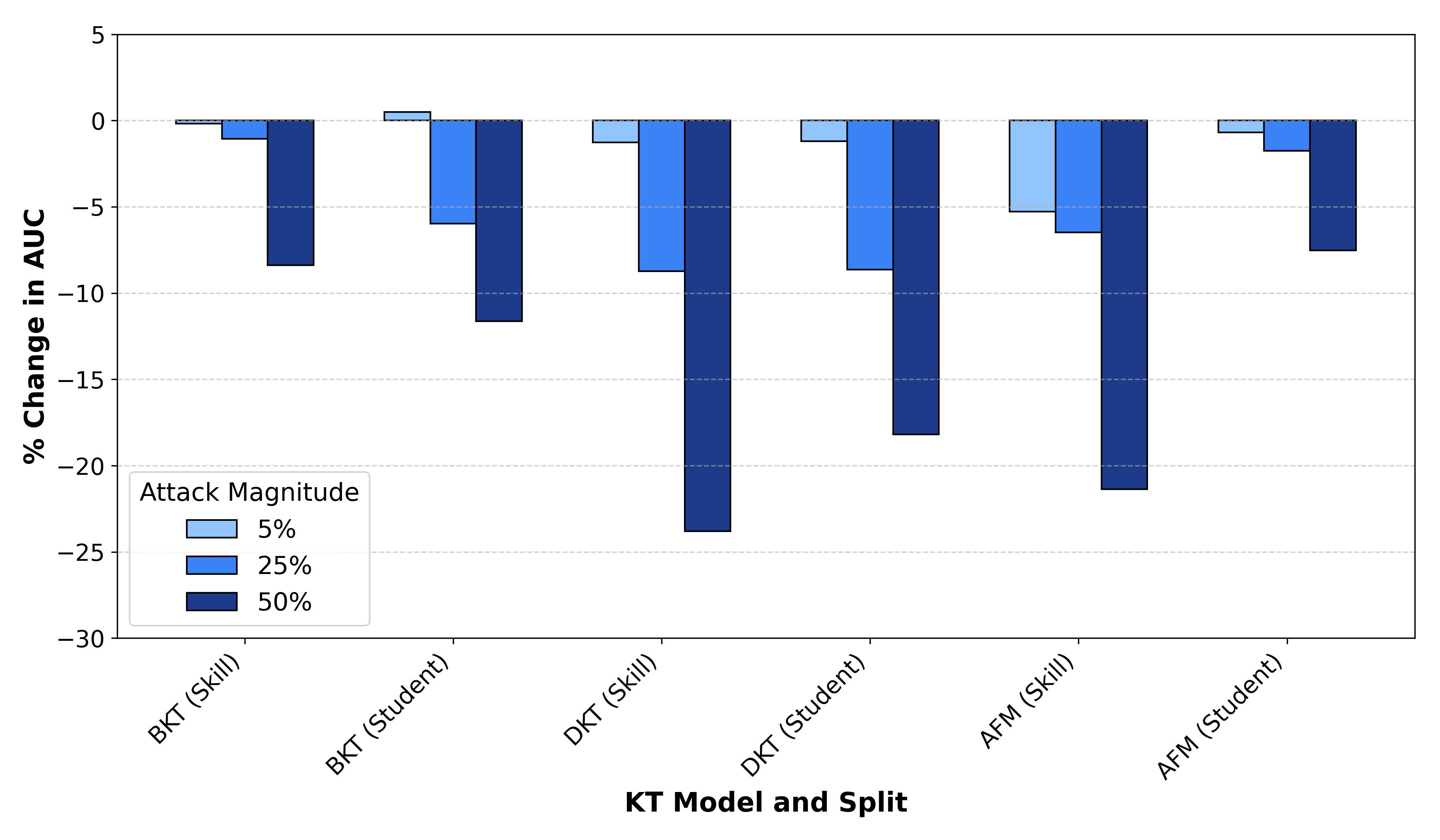}}
        \caption{Dataset A -- Random Error Attacks}
        \label{fig:assistment_random}
    \end{subfigure}
    \hfill
    \begin{subfigure}[b]{0.49\textwidth}
        \centering
        \resizebox{1.1\textwidth}{!}{\includegraphics{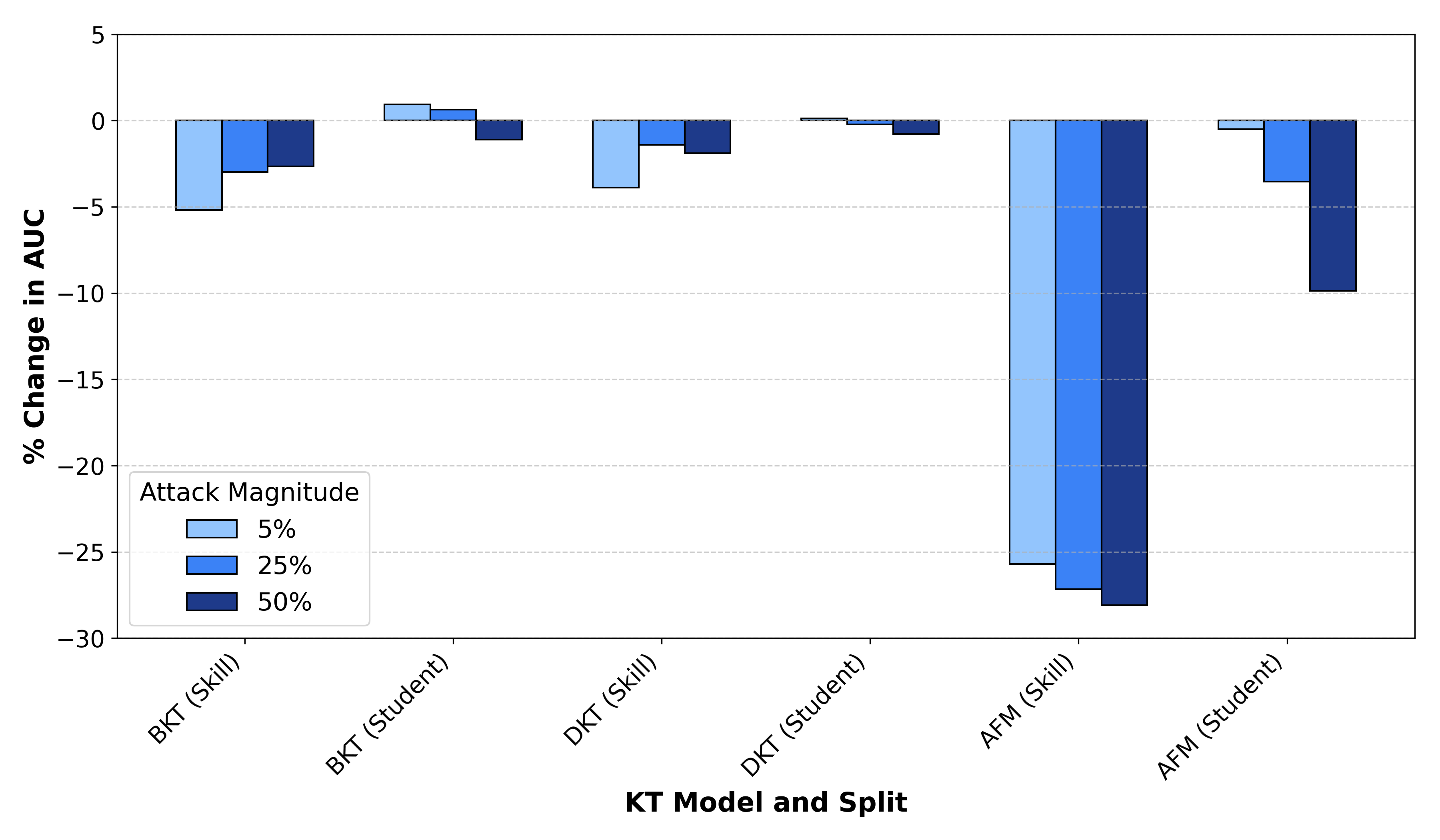}}
        \caption{Dataset B -- Random Error Attacks}
        \label{fig:hamptonalg_random}
    \end{subfigure}
    \caption{Comparison results for the Random Error Attacks across the two datasets and their two splits.}
    \label{fig:random_comparison}
\Description{Bar charts showing performance drops of most models under the experimental conditions.}
\end{figure*}

\textbf{Random Error Attacks.} As shown in Figure~\ref{fig:assistment_random} and ~\ref{fig:hamptonalg_random}, the random error attack causes substantial performance degradation for all KT models (BKT, DKT, AFM) under the skill-level split. In particular, on Dataset B, AFM drops by \(-28.08\%\) at the 50\% poisoning ratio, while on Dataset A, DKT decreases by \(-23.8\%\). In contrast, the decline under the student-level split is relatively smaller; for example, on Dataset A, AFM drops by only \(-7.54\%\) at the 50\% poisoning ratio, and DKT by \(-18.2\%\). This trend indicates that the skill-level split is more sensitive to random errors, whereas the student-level split shows a certain degree of robustness.  

From the perspective of model differences, AFM shows the highest sensitivity to Random Error (e.g., \(-28.08\%\) on Dataset B under the skill-level split). This may be related to its assumption of additive factors, where the cumulative effect of correct and incorrect answers is easily disturbed by random noise. In contrast, DKT, as a neural network model, shows a stronger performance drop on Dataset A (\(-23.8\%\)), possibly because of its higher dependency on patterns in the training data, which are disrupted by random errors.

\begin{figure*}[htbp]
    \centering
    \captionsetup{font=normalsize} 

    \begin{subfigure}[b]{0.49\textwidth}
        \centering
        \resizebox{1.1\textwidth}{!}{\includegraphics{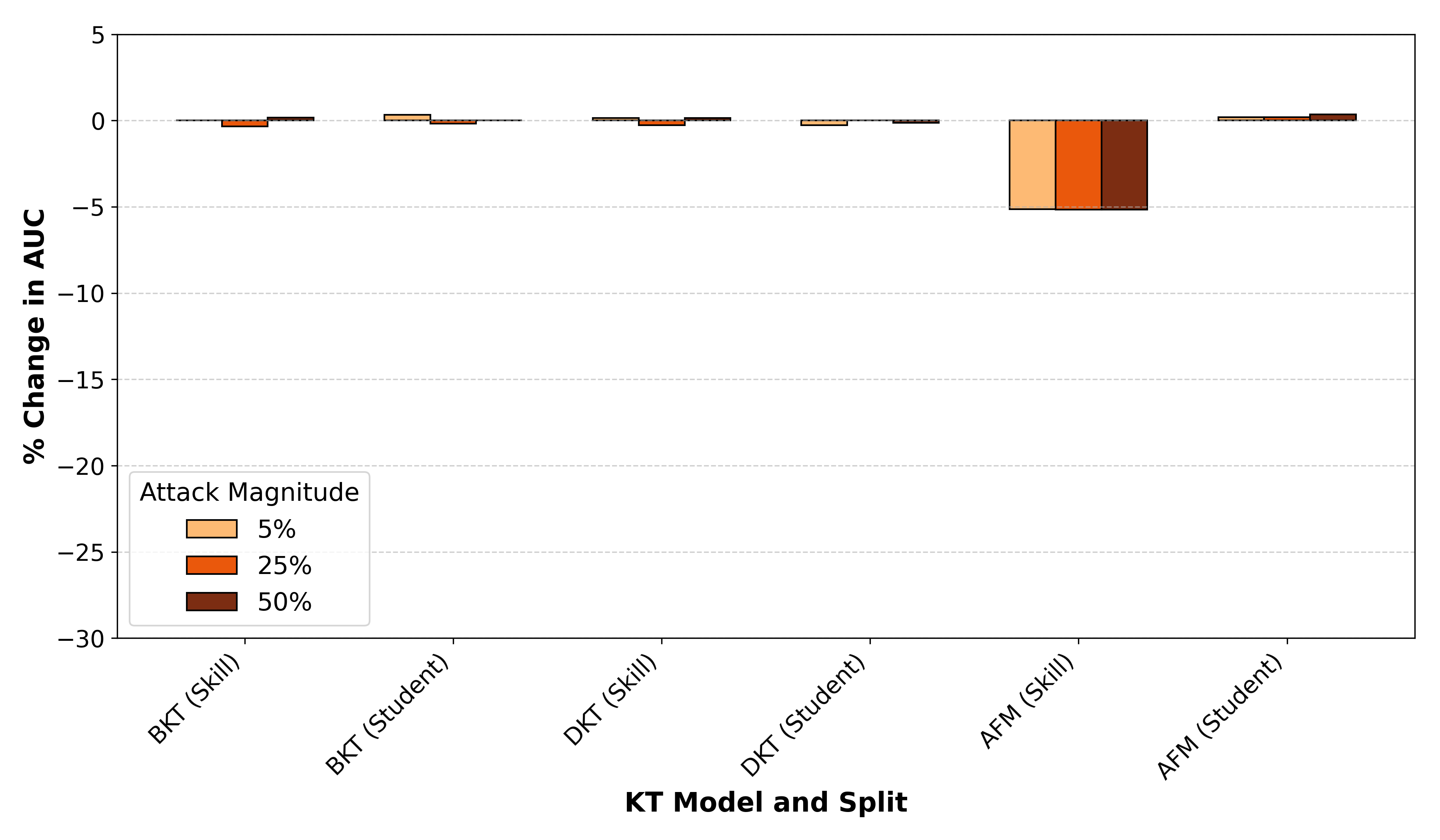}}
        \caption{Dataset A -- Sequential Pattern Attacks}
        \label{fig:assistment_seq}
    \end{subfigure}
    \hfill
    \begin{subfigure}[b]{0.49\textwidth}
        \centering
        \resizebox{1.1\textwidth}{!}{\includegraphics{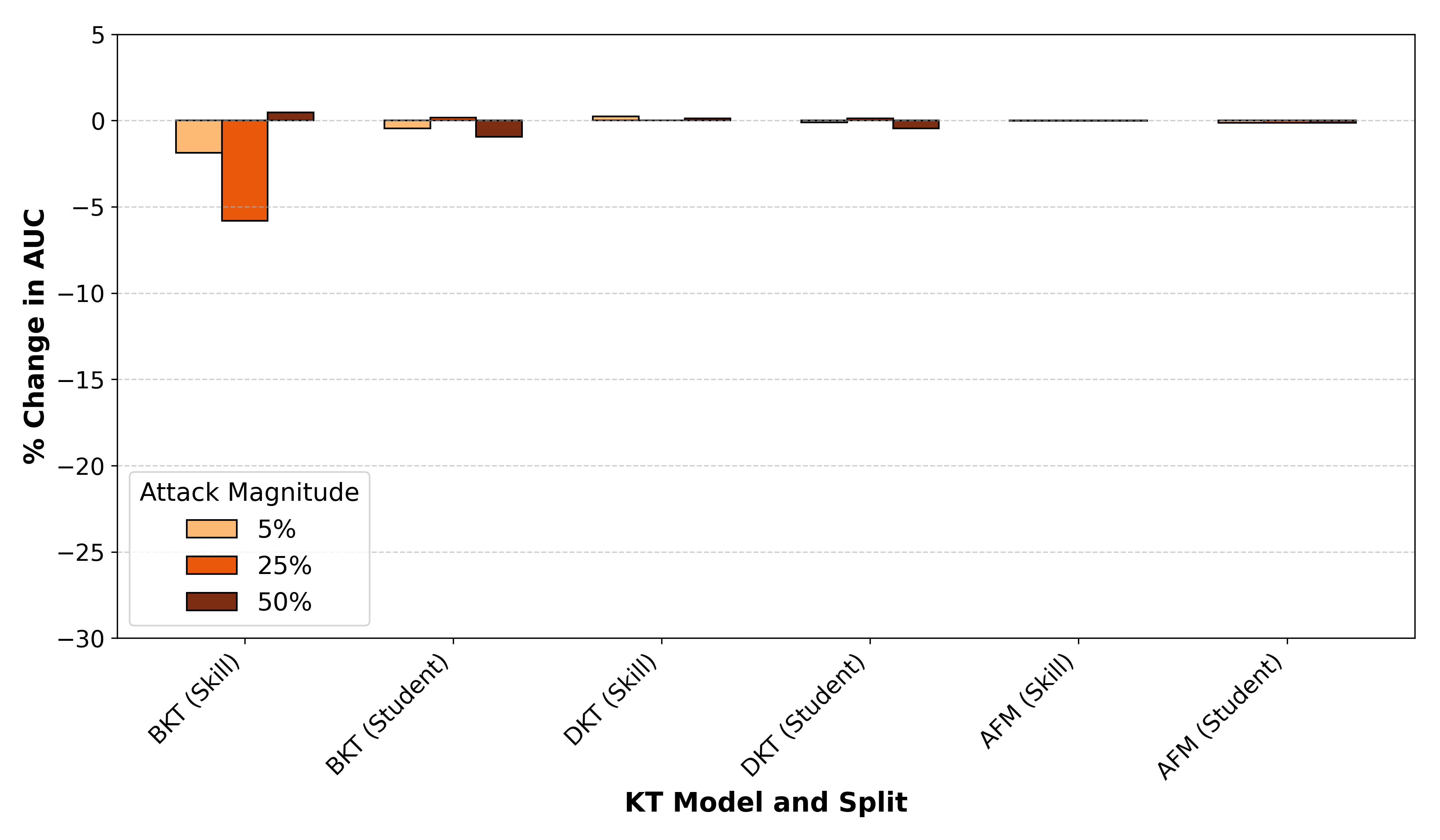}}
        \caption{Dataset B -- Sequential Pattern Attacks}
        \label{fig:hamptonalg_seq}
    \end{subfigure}
    \caption{Comparison results for the Sequential Pattern Attacks across the two datasets and their two splits.}
    \label{fig:seq_comparison}
\Description{Bar charts showing performance drops of a few models under the experimental conditions, while others remain mostly unchanged.}
\end{figure*}

\textbf{Sequential Pattern Attacks.} As shown in Figure~\ref{fig:assistment_seq} and ~\ref{fig:hamptonalg_seq}, the impact of the sequential pattern attack is relatively minor. In particular, on Dataset B, the changes at both the skill-level and student-level splits are close to \(0\%\) (e.g., AFM drops by only \(-0.01\%\) and \(-0.13\%\)). On Dataset A, AFM decreases by \(-5.17\%\) at the skill-level, while the student-level split even shows a slight positive gain (\(0.35\%\)). This indicates that the Sequential Pattern attack has only a limited impact on model performance. From the perspective of split methods, like Random Error Attack, skill-level attacks are more sensitive than student-level attacks. 

\begin{figure*}[htbp]
    \centering
    \includegraphics[width=0.7\textwidth]{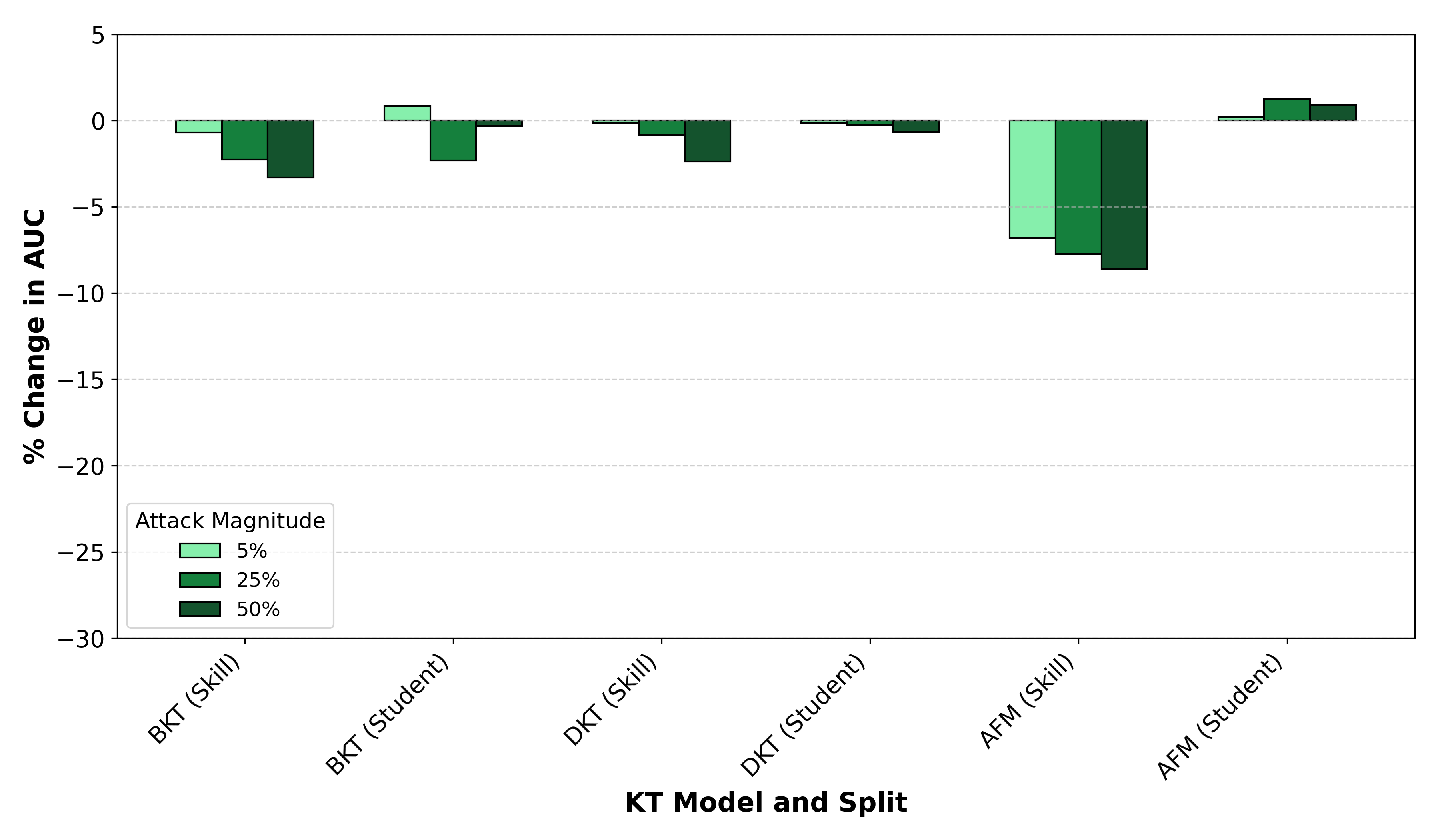}
    \caption{Dataset A -- Hint Abuse Attack. Comparison results for the Hint Abuse Attacks for the first dataset and its two splits.}
    \label{fig:assistment_hintna}
\Description{Bar charts showing slight performance drops of most models under the experimental conditions.}
\end{figure*}

\textbf{Hint Abuse Attacks.} Based on \Cref{fig:assistment_hintna}, the hint abuse attack is only applicable to Dataset A, since the ALS for Dataset B does not provide a request-hint function. The results reveal clear differences across models and data splits. Under the skill-level split, BKT shows a smaller AUC decline compared to AFM, while AFM experiences the most severe degradation, with drops of \(-6.82\%\), \(-7.74\%\), and \(-8.60\%\).  In contrast, the student-level split exhibits a different trend. BKT shifts from a positive gain of \(0.83\%\) at 5\% poisoning to decreases of \(-2.32\%\) (25\%) and \(-0.33\%\) (50\%). DKT maintains relatively minor declines (\(-0.13\%\), \(-0.27\%\), and \(-0.66\%\)), whereas AFM even shows positive gains (\(0.18\%\), \(1.23\%\), and \(0.88\%\)). This divergence suggests that the negative effects of Hint Abuse are more pronounced under the skill-level split.  

The substantial drop in AFM performance at the skill-level split (\(-8.60\%\)) likely reflects its sensitivity to frequent hint requests, which disrupt cumulative skill mastery evaluation. BKT and DKT show performance declines: BKT may interpret hints as stalled learning, while DKT’s neural architecture handles such irregular patterns more flexibly.

\section{Discussion}
\label{sec:discussion}

\subsection{Discussion of RQ1 \& RQ2}
\textbf{Reasons for the performance degradation of KT models due to gaming.}
Our results indicate that when data are contaminated by gaming behaviors, the performance of KT models is substantially impaired. This performance drop both highlights the effectiveness of DPA in replicating real-world gaming behaviors for RQ1, and exposes the limitations in the design of KT models, which reduces their reliability. There are two main reasons for the observed performance degradation: First, the assumptions underlying KT model design do not account for gaming; such behaviors violate the assumption that student learning progress depends on their engagement~\cite{koedinger2023astonishing}. The models do not consider that part of the increased engagement may be the result of gaming behaviors, which constitutes ineffective participation. Second, KT models are highly sensitive to input interaction data; thus, simulated gaming attacks that alter these inputs predictably impact model performance, consistent with previous studies~\cite{Wang2025KnowledgeTracing}.

\par From the perspective of model-specific sensitivities, the reasons for performance degradation in different models are different. Parametric models (AFM, BKT) suffer from systematic bias in global parameter estimates under large-scale label corruption (e.g., Random Errors). DKT, as a sequence-focused non-parametric model~\cite{piech2015deep}, is more robust to structured attacks (e.g., Sequential Patterns) that it can learn as temporal regularities, but vulnerable to unstructured noise that breaks sequence coherence.

\textbf{The detrimental impact of gaming behaviors on KT model performance and ALS functionality.} When student data contains a certain proportion of gaming behaviors, the KT model's quality deteriorates, impacting other services provided by the ALS to students (e.g. average AUC drop $-15.2\%$ for Random Error Attack). First, the KT model's performance degradation can lead to deviations in the ALS's personalized learning paths. The KT model's core function is to dynamically model students' knowledge based on their interaction data, thereby providing the ALS with personalized learning paths~\cite{corbett1994knowledge}. When data is contaminated by gaming behaviors, the KT model may miscalculate students' knowledge levels. For example, students who frequently abuse hints may be misjudged as having mastered a concept, leading the ALS to assign them overly complex learning materials. This not only increases the student's learning burden but also can lead to frustration and reduce motivation. Conversely, students who answer quickly and randomly may be misjudged as having insufficient knowledge, leading the ALS to assign them overly simple tasks. This overpractice, involving ineffective use of practice time beyond mastery, can hinder their learning progress by reducing exposure to more challenging material and ultimately undermines the personalized educational goals of the ALS. 

In addition, the decline in KT model performance caused by gaming behaviors will also raise fairness and ethical issues for ALS. This is reflected in two points. First, since the KT model is usually trained or updated based on the interaction data of all students, when a large amount of noise data is generated by gaming behaviors, the impact will not be limited to some gaming students, and the prediction accuracy of normal student behavior will also be reduced. The second point is that gaming students may obtain higher system scores by circumventing human efforts, while serious students’ model errors are underestimated due to data contamination. This unfairness may exacerbate the gap in the distribution of educational resources and violate the original intention of ALS to promote educational equity \cite{dai2021knowledge}. In addition, if ALS fails to effectively detect and deal with gaming behaviors, it may inadvertently encourage students to adopt similar strategies, which may undermine the integrity of the learning culture in the long run.

\textbf{Reasons why certain gaming behaviors have only a minor negative impact—or even a slight positive effect—on KT model performance.} Our results show that certain gaming behaviors, such as sequential pattern attacks, cause minimal degradation or even slight improvements in KT model performance. This can be attributed to two factors. First, KT model adaptability mitigates the impact of gaming behaviors. For instance, BKT's parameters for guess ($P(G)$) and slip ($P(S)$) partially account for gaming manifestations like sequential attacks by interpreting them as probabilistic errors rather than systematic exploitation, akin to carelessness behaviors \cite{zambrano2024investigating}; however, this can lead to model degeneracy, where noisy data inadvertently enhances fit by masking anomalies as normal learning variations~\cite{doroudi2017misidentified}. Similarly, DKT learns complex sequential patterns as standard features, and AFM’s additive-factor assumption masks systematic trial-and-error effects \cite{Effenberger2020afm}. While this reflects some robustness in KT models, it also indicates a failure to detect data contamination, as students do not progress through genuine learning as assumed. Second, a data dilution effect occurs when high proportions of gaming behaviors cause trial records to offset across skill categories, resulting in minimal AUC changes. In this effect, the aggregated noise from diverse gaming instances across students dilutes the signal of anomalous patterns, allowing the model to inadvertently treat them as balanced variations in learning trajectories. This misclassification of gaming as exploratory learning underscores the need for more sensitive monitoring mechanisms to distinguish gaming the system behaviors from authentic learning.

\textbf{Limitations and future work.} This study was limited to student- and skill-level evaluations, without examining how different DPA affect individual skills or skill groups. Such an analysis could reveal, for instance, whether harder skills are more vulnerable to gaming~\cite{Huang2023GamingDetection}. How KT model robustness varies across skill domains represents a valuable direction for future work. Moreover, our evaluation focused primarily on overall AUC; future work can investigate how poisoning distorts learner-state trajectories, mastery estimates, and downstream recommendations. Additionally, our experiments only used math-domain datasets. Therefore, broader domain evaluation is needed to assess generalization. Finally, while our use of DPA provides a controlled method to simulate gaming, future research must thoughtfully address the ethical implications of applying adversarial frameworks in educational contexts. Key considerations include ensuring transparency, preventing the misuse of such techniques, and aligning detection and intervention mechanisms with supportive—rather than purely punitive—pedagogical principles.

\subsection{Discussion of RQ3}
RQ3 investigates how DPA detection methods can offer new perspectives for gaming detection. Traditional gaming detectors mainly rely on knowledge-engineering models and supervised machine learning. While effective, these approaches suffer from label dependency and poor generalization \cite{Huang2023GamingDetection}. DPA detection methods, which often use unsupervised techniques to identify data contamination by simulating gaming behaviors, provide the following insights:

1. \textbf{Reduced labeling costs:} Traditional gaming detection is largely supervised, requiring extensive human-labeled data to train features such as help-seeking or error patterns. This increases cost and limits applicability to new datasets~\cite{paquette2019comparing}. In contrast, DPA detection uses unsupervised methods that rely solely on intrinsic data anomalies (e.g., extreme deviations in statistical metrics) to identify gaming patterns without labels. This offers a low-cost expansion pathway for traditional detectors, particularly in educational settings where labeled data is scarce. It can serve as a pre-screening tool to reduce the need for fully supervised training.

2. \textbf{Greater generalizability:} Traditional machine learning models often suffer substantial performance drops, or even fail entirely on new systems or datasets, as they heavily depend on the specific distribution of training data and previously learned labels \cite{Huang2023GamingDetection}. Unsupervised DPA detection methods, by contrast, do not rely on pre-labeled gaming data. Instead, they detect anomalous behaviors through statistical outliers, clustering, or pattern discovery. By focusing on the intrinsic structure and distribution of the data rather than fitting specific training labels, these methods are more likely to generalize across systems and datasets.

3. \textbf{Ease of integration with defensive strategies:} Unsupervised detection methods can be seamlessly incorporated into data cleaning workflows \cite{Fan2022SurveyDataPoisoning}, proactively removing anomalous or potentially gaming-related samples before training KT models. This is not only a defensive measure but also a proactive “turn defense into offense” strategy: by cleaning the data upfront, models are less likely to be poisoned or manipulated during training, mitigating the impact of future gaming the system behaviors. Analogous to sanitization defenses in machine learning security, this cleaning step can enhance model robustness against anomalous behaviors while reducing reliance on specific patterns. Moreover, datasets preprocessed with unsupervised detection provide a healthy baseline for subsequent monitoring, anomaly detection, and behavior strategy optimization, achieving a closed loop of prevention and defense.

Despite these advantages, applying DPA detection to gaming the system behavior also entails potential drawbacks, such as higher computational costs and possibly elevated false positives during early iterations. Nevertheless, considering its many potential benefits, we encourage future research to explore this approach.

\subsection{Implications of This Study}

Our study carries implications in four key areas detailed below.

\begin{itemize}
    \item \textbf{Robust KT model design and gaming behaviors detection.} 
    Gaming behaviors can degrade KT model performance and violate the assumption that students gradually master skills through repeated practice, allowing KT models to operate under invalid assumptions. Developing robust detection mechanisms is critical to identify gaming behaviors and protect the overall performance, reliability, and fairness of ALS. Future KT model designs could also integrate noise-filtering mechanisms to reduce sensitivity to anomalous or contaminated data.

    \item \textbf{Optimizing ALS system design and educational interventions.} 
    ALS systems can embed real-time data quality monitoring to dynamically detect gaming behaviors and adjust recommendation strategies. Recommendations for suspected gaming behaviors can be paused, while resources are prioritized for students with normal behavior, preventing contamination from affecting other ALS algorithms. Furthermore, long-term presence of gaming behaviors may undermine academic integrity. Educators can leverage detection results to implement targeted interventions, helping students develop proper learning habits.
    
    \item \textbf{Cross-disciplinary insights from DPA detection.} 
    Methods developed for detecting DPA offer insights for gaming behaviors detection. By adapting DPA detection techniques, traditional gaming behaviors detectors can be improved or new unsupervised detectors can be developed. This approach not only prevents educational dataset contamination but also ensures that KT model assumptions are upheld and that ALS maintains fairness.

    \item \textbf{Integrating LA and adversarial machine learning holds potential.} 
    By combining KT models with DPA, we tested the robustness of KT models under large-scale gaming behaviors, reaffirming their sensitivity to input data quality. This characteristic is also observed in other LA models, such as learning analytics dashboards and student behavior prediction systems \cite{Liu2024}, highlighting the potential of adversarial approaches for evaluating and enhancing model robustness. Other LA models can leverage adversarial machine learning techniques to develop more resilient LA systems, thereby promoting personalized and equitable data-driven education.
\end{itemize}

\section{Conclusion}
This study introduced a novel simulation-based approach to quantify the impact of student gaming behaviors on knowledge tracing models. By framing gaming as a form of data poisoning, we employed DPA to generate controllable, adversarial interactions that mimic key behavioral patterns (e.g., random guessing, hint abuse, systematic trialing). This approach allowed us to systematically evaluate how such data contamination degrades the performance of KT models (BKT, AFM, DKT) under varying attack types and intensities (RQ1 \& RQ2). Our results confirm that gaming behaviors can substantially impair model accuracy and violate core KT assumptions, potentially compromising the personalized pathways generated by adaptive learning systems. Furthermore, we show how detection methods from adversarial machine learning can inform the development of more robust, less label-dependent gaming detectors (RQ3). We encourage learning analytics researchers and practitioners to adopt adversarial simulation techniques to proactively assess and strengthen the robustness of educational models against data quality threats.

\begin{acks}
The work of Valdemar Švábenský on this paper was supported by the Czech Science Foundation (GAČR) grant no. 25-15839I.
We thank the authors of the research datasets (\cite{Patikorn2020datasetASSISTments} and \cite{Baker2008datasetHamptonAlg}) for sharing their data.
We also thank Radek Pelánek for providing useful ideas on drafts of this paper.
\end{acks}

\bibliographystyle{ACM-Reference-Format}
\bibliography{main}

\end{document}